\documentclass[sigconf]{acmart}
\usepackage{multirow}
\usepackage{balance}

\def\BibTeX{{\rm B\kern-.05em{\sc i\kern-.025em b}\kern-.08emT\kern-.1667em\lower.7ex\hbox{E}\kern-.125emX}}
    
\copyrightyear{2019}
\acmYear{2019}
\setcopyright{iw3c2w3}
\acmConference[WWW '19 Companion]{Companion Proceedings of the 2019 World Wide Web Conference}{May 13--17, 2019}{San Francisco, CA, USA}
\acmBooktitle{Companion Proceedings of the 2019 World Wide Web Conference (WWW '19 Companion), May 13--17, 2019, San Francisco, CA, USA}
\acmPrice{}
\acmDOI{10.1145/3308560.3316757}
\acmISBN{978-1-4503-6675-5/19/05}

\begin{document}

%
\title{A graph-structured dataset for Wikipedia research}

%
\author{Nicolas Aspert, Volodymyr Miz, Benjamin Ricaud, and Pierre Vandergheynst}
\affiliation{\institution{LTS2, EPFL, Station 11, CH-1015 Lausanne, Switzerland}}
\email{firstname.lastname@epfl.ch}

%
\renewcommand{\shortauthors}{Aspert, et al.}

%
\begin{abstract}
Wikipedia is a rich and invaluable source of information. Its central place on the Web makes it a particularly interesting object of study for scientists. Researchers from different domains used various complex datasets related to Wikipedia to study language, social behavior, knowledge organization, and network theory. While being a scientific treasure, the large size of the dataset hinders pre-processing and may be a challenging obstacle for potential new studies. This issue is particularly acute in scientific domains where researchers may not be technically and data processing savvy. On one hand, the size of Wikipedia dumps is large. It makes the parsing and extraction of relevant information cumbersome. On the other hand, the API is straightforward to use but restricted to a relatively small number of requests. The middle ground is at the mesoscopic scale, when researchers need a subset of Wikipedia ranging from thousands to hundreds of thousands of pages but there exists no efficient solution at this scale.

In this work, we propose an efficient data structure to make requests and access subnetworks of Wikipedia pages and categories. We provide convenient tools for accessing and filtering viewership statistics or "pagecounts" of Wikipedia web pages. The dataset organization leverages principles of graph databases that allows rapid and intuitive access to subgraphs of Wikipedia articles and categories. The dataset and deployment guidelines are available on the LTS2 website \url{https://lts2.epfl.ch/Datasets/Wikipedia/}.
\end{abstract}

%
%

%

%

%
\maketitle

\section{Introduction}

Wikipedia is one of the most visited websites in the world. Millions of people use it every day searching for answers to various questions ranging from biographies of popular figures to definitions of complex scientific concepts. As any other website on the Web, Wikipedia stores web logs that contain viewership statistics of every page. Worldwide popularity of this free encyclopedia makes these records an invaluable resource of data for the research community. In this work, we present a convenient and intuitive graph-based toolset for researchers that will ease the access to this data and its further analysis.

Wikimedia Foundation, the organization that hosts Wikipedia, makes the web activity records and the hyperlinks structure of Wikipedia publicly available so anyone can access the records either through an API or through the database dump files. Even though the data is well structured, efficient pre-processing and wrangling requires data engineering skills. First, the dumps are very large and it takes a long time for researchers to load and filter them to get what they need to study a particular question. Second, although the API is well documented and easy to use, the number of queries and the response size are very limited. 

\begin{figure}
  \includegraphics[width=8.5cm, trim={1.5cm 0cm 0cm 0cm}, clip]{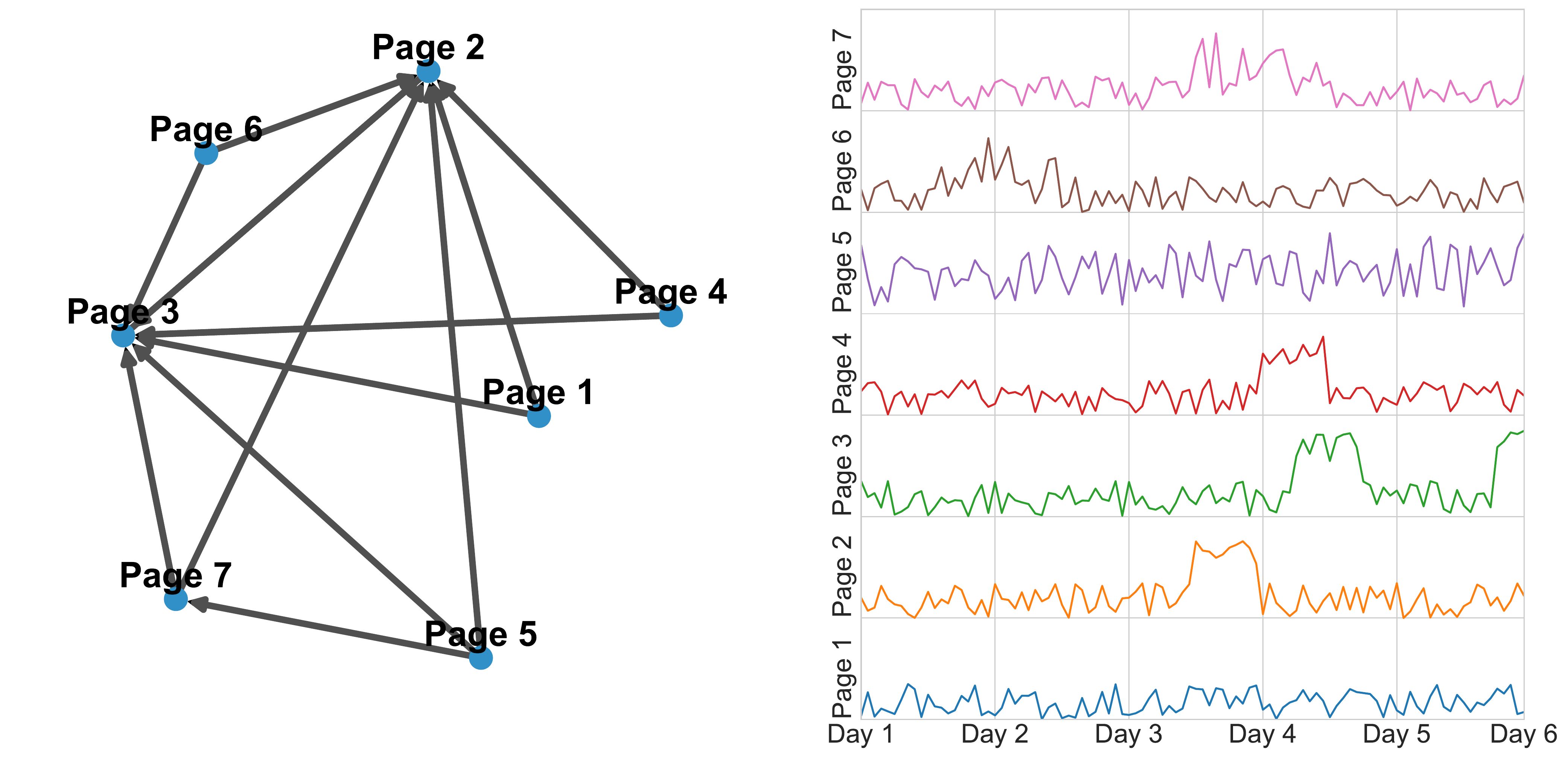}
  \caption{A subset of Wikipedia web pages with viewership activity (pagecounts). \textbf{Left:} Wikipedia hyperlinks network, where nodes correspond to Wikipedia articles and edges represent hyperlinks between the articles. \textbf{Right:} hourly page-view statistics of Wikipedia articles (right).}
  \label{fig:teaser}
\end{figure}

Even though the API is quite convenient, it can cause reproducibility issues. The network of hyperlinks evolves with time so the API can only provide the latest network configuration. To solve this problem, as a workaround, researchers use static pre-processed datasets. Two of the most popular datasets for Wikipedia network research are available on the SNAP archive, Wikipedia Network of Hyperlinks~\cite{SNAPwikihypernet} and Wikipedia Network of Top Categories~\cite{SNAPtopcat,klymko2014using, yin2017local}. The initial publications referring to these datasets have been cited more than 1000 times, showing the high interest in these datasets. These archives were created from Wikipedia dumps in 2011 and 2013 respectively. However, Wikipedia has evolved since then and Wikipedia research community would benefit from being able to access more recent data.

Multiple studies have analyzed Wikipedia from a network science perspective and have used its network structure to improve Wikipedia itself or to gain insights into collective behavior of its users. In~\cite{zesch2007analysis}, Zesch and Gurevych used Wikipedia category graph as a natural language processing resource. Buriol et al. \cite{buriol2006temporal} studied the temporal evolution of Wikipedia hyperlinks graph. Bellomi and Bonato conducted a study \cite{bellomi2005network} of macro-structure of English Wikipedia network and cultural biases related to specific topics. West et al. proposed an approach enabling the identification of missing hyperlinks in Wikipedia to improve the navigation experience~\cite{west2015mining}.

Another direction of Wikipedia research focuses on the pagecounts analysis. Moat et al. \cite{moat2013quantifying} used Wikipedia viewership statistics to gain insights into stock markets. Yasseri et al. \cite{yasseri2012dynamics} studied editorial wars in Wikipedia analyzing activity patterns in viewership dynamics of articles that describe controversial topics. Mesty\'an et al. \cite{mestyan2013early} demonstrated that Wikipedia pagecounts can be used to predict the popularity of a movie. Collective memory phenomenon was studied in~\cite{garcia2017memory}, where authors analyzed visitors activity to evaluate the reaction of Wikipedia users on aircraft incidents.

The hyperlink network structure, on one hand, and the viewership statistics (pagecounts) of Wikipedia articles, on the other hand, have attracted significant attention from the research community. Recent studies open new directions where these two datasets are combined. The emerging field of spatio-temporal data mining \cite{atluri2018spatio} highlighted an increasing interest and a need for reproducible network datasets that contain dynamically changing components. Miz et al~\cite{miz2019wikipedia} adopted an anomaly detection approach on graphs to analyze the visitors' activity in relation to real-world events. 

Following the recent advances of scientific research on Wikipedia, in this work, we focus on two components: the \textit{spatial} component (Wikipedia hyperlinks network) and the \textit{temporal} component (pagecounts). We design a database that allows querying this hybrid data structure conveniently (see Fig.~\ref{fig:teaser}). Since Wikipedia web logs are continuously updating, we designed this database in a way that will make its maintenance as easy and fast as possible. 



\section{Dataset}

There are multiple ways to access Wikipedia data but none of them provide native support of a graph data structure. Therefore, if researchers want to study Wikipedia from the network science perspective, they have to create the graph themselves, which is usually very time-consuming. To do that, they need to pre-process large dumps of data or to use the limited API.

In spatio-temporal data mining \cite{atluri2018spatio}, researchers are most interested in the dynamics of the networks. Hence, when it comes to Wikipedia analysis, one needs to merge the hyperlinks network with page-view statistics of the web pages. This is another large chunk of data, which requires another round of time-consuming pre-processing.

After the pre-processing and the merge is completed, researchers usually realize that they do not need the full network and the entire history of visitors' activity. However, there is no easy workaround: in order to get a certain subset of pages for a specified period, everyone has to perform the aforementioned steps.

In this paper, we propose a graph-based solution that eliminates the pre-processing steps described above. We present a graph database that simplifies access to Wikipedia data dumps and its viewership statistics. With a set of intuitive queries, we provide the following features:

\begin{itemize}
\item Get relatively large subgraphs of Wikipedia pages (1K--100K nodes) without redirects.
\item Use filters by the number of page views, category/sub-category, graph measures (n-hop neighborhood of a node, node degree, page rank, centrality measures, and others).
\item Get viewership statistics for a subset/subgraph of Wikipedia pages.
\item Get a subgraph of pages with a number of visits higher than a threshold, in a predefined range of dates.
\end{itemize}

The database allows its users to return subgraphs with millions of links. However, requesting a large subgraph from the database may take several hours. Besides, it may require a large amount of memory on the hosting server. Such queries may cause an overload of the database server that has to process queries from multiple users at the same time. Therefore, instead of setting up a remote database server, we have decided to provide the code to deploy a local or cloud-based one from Wikipedia dumps. This will allow researchers to explore the dataset on their own server, create new queries and, possibly, contribute to the project.

Lastly, the database will be updated every month and will be consistent with the latest Wikipedia dumps. This gives the researchers the ability to reproduce previous studies on Wikipedia data and to conduct new experiments on the latest data.
The dataset and the deployment instructions are available online~\cite{LTS2wikidataset}.

\section{Framework}

\subsection{Graph structure}
\begin{figure}
  \includegraphics[width=7cm]{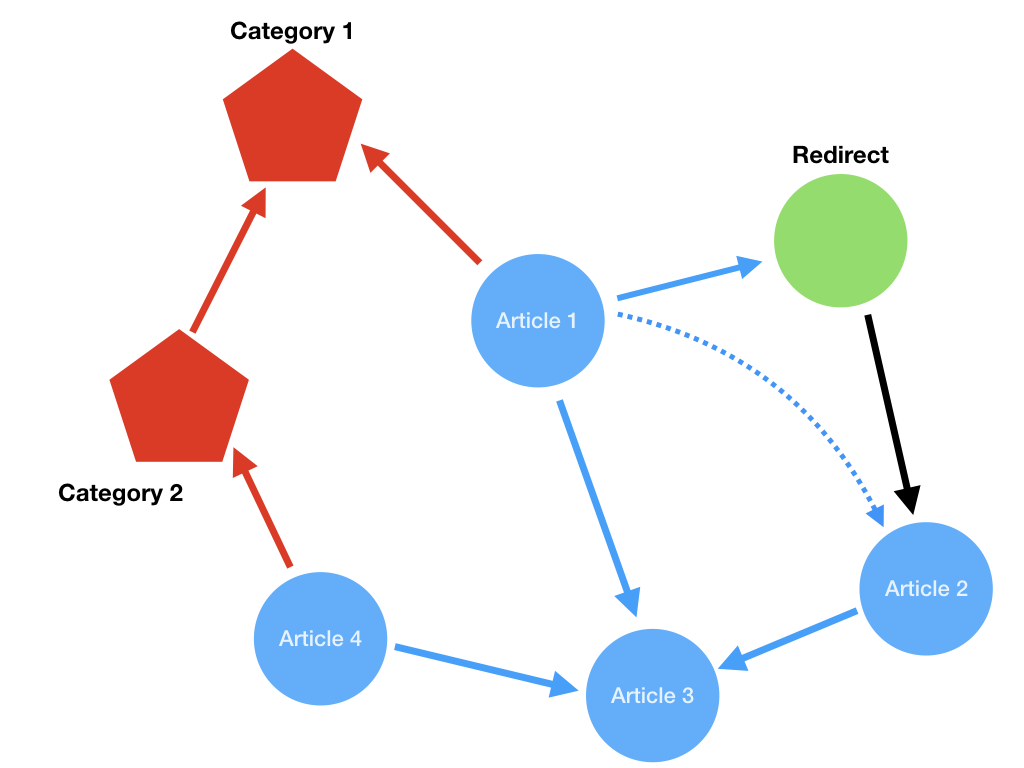}
  \caption{Wikipedia graph structure. In blue: articles and hyperlinks referring to them. In red: category pages and hyperlinks connecting the pages or subcategories to parent categories. In green: a redirected article, i.e. Article 1 refers to Article 2 via the redirected page. In black: a redirection link. The blue, dashed line, is the new link created from the redirection.}
  \label{fig:graphstructure}
\end{figure}

Wikipedia network of articles is stored in a property graph database. The graph is a multigraph with different kinds of nodes and links. The different objects are described in Table~\ref{tab:entities} and on Fig.~\ref{fig:graphstructure}.
\begin{table*}[ht]
\begin{center}
\caption{Entities in the graph\label{tab:entities}}
{\small 
\begin{tabular}{ccc}
\toprule
{\bf Name} & {\bf Nature} & {\bf Description} \\ 
\midrule
article & node & Wikipedia article\\ 
category & node & Wikipedia category article\\ \hline
links\_to & link & hyperlink between 2 articles\\ 
belongs\_to & link & hyperlink between an article or subcategory and a category page \\ 
\bottomrule
\end{tabular}
}
\end{center}
\end{table*}
Wikipedia articles are nodes of the graph and the hyperlinks pointing to different pages are recorded as directed edges between the nodes. The categories of Wikipedia articles are structured as a graph as well. Indeed, in the encyclopedia, categories are pages with a textual description, and they refer to their elements (articles or subcategories) with hyperlinks. Hyperlinks are present in each article pointing to the categories they belong to. Inside the graph database, articles and category pages are distinguished as nodes of different nature (different labels). 

In order to navigate between articles and categories conveniently, we introduce two types of links. The ''links\_to`` relations are hyperlinks between articles (excluding categories), and the ''belongs\_to`` relations are linking articles to their categories or subcategories to their parent categories. These latter edges are built from the hyperlinks within pages pointing to category pages.

\textbf{Graph structure.} The category structure is shaped as a tree. The advantage of such structure is that it is easy to handle it when the user creates articles and wants to classify them in subcategories. However, it makes it difficult to retrieve the set of all articles belonging to a given category (or subcategory). One has to explore all the hierarchy of subcategories within it and collect all the encountered articles. Therefore, we choose the graph database structure to simplify this task. Traversing and performing the breadth-first search in the graph is one of the basic functions of a graph database making this solution a more efficient alternative.

\textbf{Redirects.} In order to handle renamed or merged pages, Wikipedia relies on a redirection approach. When renaming a page, moderators create a new page with a new title but they do not remove the initial page in order for the hyperlinks from articles pointing to remain valid. Instead, in order to avoid this, the initial page becomes a "redirect", a page that automatically redirects a visitor to a new page. Redirect pages are invisible to users. We removed these redirection pages from our dataset by redirecting the hyperlinks pointing to the correct article (the blue dashed arrow of Fig.~\ref{fig:graphstructure}). First, it simplifies the queries when exploring the graph. Second, it makes it easier for users to understand the structure. Lastly, it halves the number of nodes in the graph: at the time of writing, the number of articles in the English Wikipedia is close to 6 million while the number of redirects is around 8 million.  


\subsection{Time series of visits}\label{sec:timeseries}

The time series of visits are stored separately in a NoSQL database in the form of a collection of indexed \textit{key:value} pairs. Each key is a pair \textit{(page id, time-stamp)} and the value is the \textit{number of visits} during the hour given by the time-stamp for the page associated to the page id.

This structure provides a flexible way of recording new entries following the evolution of time, the page creation and deletion that occur in the encyclopedia. Querying a specific period of time is very convenient and efficient as well. It is done by submitting a request with a specific range of key values (a range applied to the time-stamp key of the key couple).

In order to reduce the amount of data to be stored, we introduce a threshold for a number of visits per page per day. We store the number of hourly visits for an article if the daily total of its visits is above this threshold (100 by default). This reduces the number of entries by an order of magnitude without losing relevant information. The database handles missing records automatically, which is also very convenient.

\subsection{Data extraction and pre-processing}

Before creating the database, we perform the following pre-processing steps. After having downloaded Wikipedia dumps~\cite{wikidumps}, we parse the SQL files to extract the titles of articles and categories, page and category ids, and the hyperlinks. Before storing the data in the graph database, we remove the redirects and modify the hyperlinks pointing to them to link to the correct articles. After these steps are completed, we load the data into the graph database.

We download the pagecounts dumps~\cite{wikidumpsvisits} (number of visits per page per hour), and extract the hourly visits. As stated in section~\ref{sec:timeseries}, we remove entries with a low number of visits. If a page has less than 100 daily visits, we do not store visit records for that page and that day. We store the values above this daily threshold in the NoSQL database with an hourly resolution.



\subsection{Database update}

The separation of the graph and time series data simplifies the update process and the maintenance. We update every part separately and in a different manner.

We update the graph database periodically. At the time of writing, we do it on a monthly basis. The update follows monthly releases of Wikipedia dumps. Every month, we compare the new and the previous version dumps of pages and links. We add the new nodes and links to the database and delete the removed ones.

We perform daily updates of the time series database. Every day, we add 24 new entries (one per hour) for each article (only those entries that surpass the 100 daily visits threshold). 

\begin{table}[ht!]
\begin{center}
\caption{Number of nodes in the subgraph of the ''Physics`` category\label{tab:physics_cat}}
\begin{tabular}{crr}
\toprule
{\bf Depth} & {\bf Articles} & {\bf Subcategories} \\ 
\midrule
1 & 69 & 27 \\ 
2 & 2'263 & 206 \\ 
3 & 10'128 & 970 \\ 
4 & 33'917 & 3'711 \\ 
5 & 80'349 & 16'917 \\ 
6 & 232'818 & 74'004 \\ 
7 & 2'041'232 & 251'551 \\ 
\bottomrule
\end{tabular}
\end{center}
\end{table}

\section{Performance of the queries}
We constructed the graph of English Wikipedia pages based on the August 1st 2018 SQL dumps. After resolving redirects, the graph consists of about ca. 7.4 million articles, comprising both regular pages (ca. 5.7 million) and category pages (ca. 1.7 million), and ca. 511 million edges. Once the data was imported into the graph database, we ran queries to extract various sub-graphs, e.g. retrieve all pages and subcategories belonging to a given category and all the links between these pages. Table~\ref{tab:performances} demonstrates query results and the time required to process them. The presented results have been computed on a 24 cores Intel Xeon E5 system, equipped with SSD drives and using the Neo4j open-source database. Given the highly connected structure of Wikipedia, we had to restrict the depth of certain queries, as the returned set expands dramatically. 
\begin{table*}[ht]
\begin{center}
\caption{Size and performances for different subgraph requests\label{tab:performances}}
\begin{tabular}{crrrcc}
\toprule
{\bf Category} & {\bf Articles} & {\bf Hyperlinks}  & {\bf Subcategories} & {\bf Search depth} & {\bf Processing time}\\ 
\midrule
 & 571 & 5'165 & 202 & 2 & 0.4 s \\ 
Philosophy & 5'370 & 177'754 & 1'144 & 3 & 29.7 s \\ 
 &  26'480 & 1'094'550  & 4'084  & 4 & 574 s \\ \hline
 & 2'263 & 27'911 & 207 & 2 & 3.3 s \\ 
Physics & 10'128 & 223'870 & 971 & 3 & 55 s \\ 
 & 33'917 & 972'206 & 3'712 & 4 & 501 s \\ \hline
\multirow{2}{*}{Science} & 1'762 & 19'189 & 455 & 2 & 3 s\\ 
 & 18'751 & 260'043 & 2'842 & 3 & 292 s\\ \hline
\multirow{2}{*}{Actors} & 1'107 & 3'313  & 654 & 2 & 1.6 s \\ 
 & 10'805 & 47'196  & 2'922 & 3 & 90 s \\ \hline
 & 859 & 6'598 & 223 & 2 & 1 s \\ 
Global conflicts & 6'179 & 152'517 & 1'208 & 3 & 48.5 s \\ 
 & 22'663 & 706'357 & 3'905 & 4 & 541 s \\ \hline
Exoplanets & 989 & 18'926 & 69 & unlimited & 0.8 s \\ 
\bottomrule
\end{tabular}
\end{center}
\end{table*}

While it is possible to use traditional relational databases to store the pages and links information, retrieving a subgraph using such a structure would require an increasing number of subqueries when increasing the depth of the subgraph queried, resulting in longer processing time and complex query syntax. The subgraph requires multiple queries to find all the nodes belonging to the subgraph, then an additional search to find all the edges connecting any of the nodes in the set. Experiments have been conducted using direct processing of pre-processed page and link data using Apache Spark, and also using a relational database (PostgreSQL). Data used to perform the databases comparison is based on a trimmed down version of the Wikipedia SQL dumps, with additional processing to replace the target of links, expressed as page title and namespace, by the page unique id, and removing redirects. This pre-processing, combined with database indexes, leads to simpler and faster queries. 

Queries performed on a database are often at least an order of magnitude faster than direct queries on the raw page and links data using Apache Spark. For instance, querying the sub-category graph of the ''Physics`` category with a depth 2 requires approximately 5 minutes to retrieve all the nodes belonging to the subgraph, and also several additional minutes to retrieve its edges, whereas the same data is completely extracted in less than 5 seconds using the graph database. Using a relational database improves the situation, as the nodes of the subgraph are returned in less than a second. Returning the edges from the subgraph remains however time-consuming (ca. 10 to 40 seconds in our experiments), in addition to requiring multiple nested queries whose complexity increases with the search depth. In that particular example, given the relatively small size of the result, timings can be heavily impacted by the cache of each application, as well as by the system they run on, especially by the presence of SSD vs. HDD. For instance, using a database query (on the graph database or the relational one) to retrieve a subgraph of depth 3, then retrieving the same subgraph of depth 2 will most likely only use the cache and yield much faster results. When the search depth increases sufficiently, the relational database can lead to faster processing than the graph database, at the expense of query complexity.

Similarly, we queried subgraphs consisting of page neighbors (i.e. connected via a ''links\_to`` relation), up to a certain depth. We also restricted the queries by the number of outgoing links from the top page since some of them have a huge number of direct connections. We provide the results of these queries in Table~\ref{tab:perf_links}. Increasing the depth of such queries (e.g. for depth greater than one) leads to the large responses, resulting in long processing time (cf. the ''Computer science`` entry in Table~\ref{tab:perf_links}).

\begin{table*}[ht]
\begin{center}
\caption{Size and performances for article neighbor subgraph requests\label{tab:perf_links}}
\begin{tabular}{crrrcc}
\toprule
{\bf Page} & {\bf Articles} & {\bf Hyperlinks}  & {\bf Subcategories} & {\bf Search depth} & {\bf Processing time}\\ 
\midrule
Switzerland & 1'400 & 144'911 & 24 & 1 & 4.5 s \\ \hline
United States & 2'215 & 258'939 & 28 & 1 & 17.5 s \\ \hline
Charlie Chaplin & 1'289 & 147'203 & 23 & 1 & 4 s \\ \hline
Albert Einstein & 1'025 & 114'518 & 30 & 1 & 2.3 s \\ \hline
\multirow{2}{*}{Computer science} & 684 & 47'067 & 13 & 1 & 1 s \\ 
 & 68'756 & 7'883'471 & 1'450 & 2 & 3'600 s \\ 
\bottomrule
\end{tabular}
\end{center}
\end{table*}

\section{Use Cases and Applications}\label{sec:usecases}

\subsection{Subgraphs of categories}

Wikipedia articles are classified according to the category hierarchy established by the contributors. The absence of strict guidelines or strong authority on the category labeling has led to a complex category schema. Gathering all the pages belonging to a category is a difficult task at the moment. It requires visiting all the subcategories belonging to the initial category and collecting the articles they refer to. Furthermore, the absence of dedicated curation leads to the collection of several subcategories (and hence articles) only remotely related to the original one. Those sub-categories can be very generic and encompass a large number of articles, e.g. one of the sub-categories of ''Physics`` is ''Writing systems`` (linked via ''Physical systems``). Indeed, the very deep hierarchy and the lack of tools for accessing the network of categories make it impossible to have a global view on the structure and efficient maintenance. 

\begin{figure}
  \includegraphics[width=7cm]{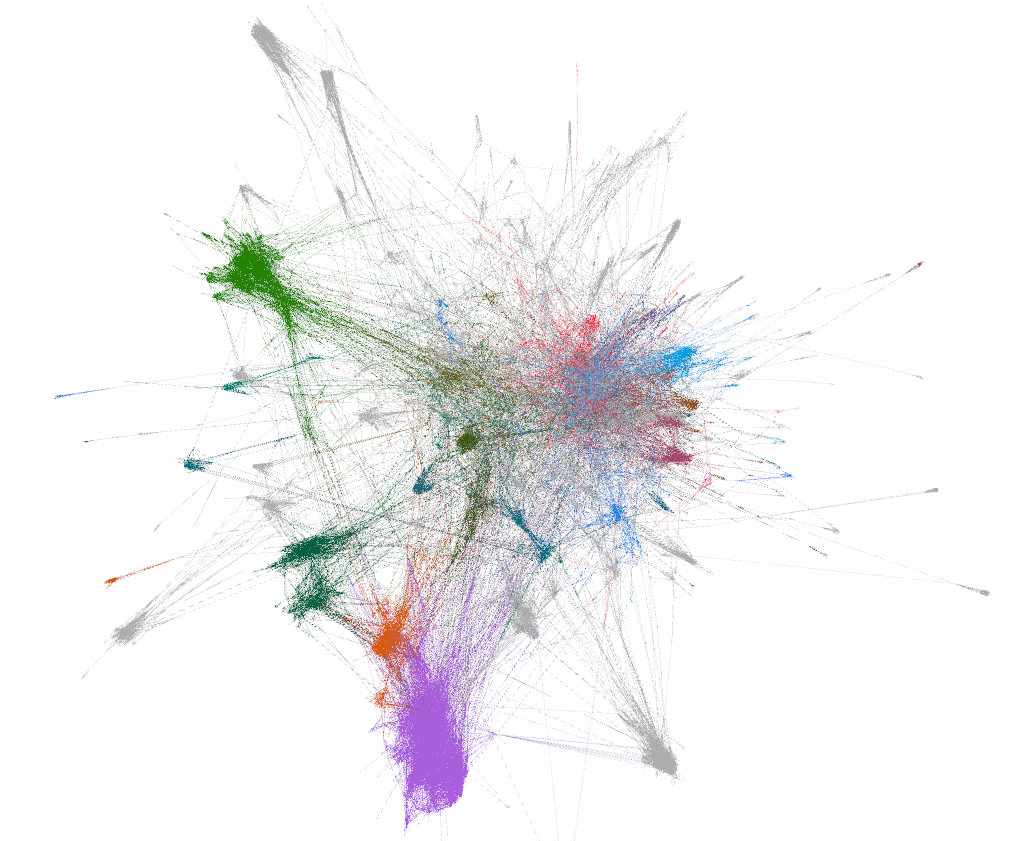}
  \caption{Network view of a reduced subset of Wikipedia web pages ($\sim$20K nodes, $\sim$100K edges) using the method described in~\cite{miz2019wikipedia}. Nodes correspond to popular articles with spikes of visits during the period Oct. 2014 - Apr. 2015. They are connected by links with strength related to the correlation of their viewership activity. Real-world events trigger spikes in the number of visits in groups of pages, forming the clusters in the network.}
  \label{fig:wiki_snapshot}
\end{figure}

To illustrate the complexity of the category structure, we run multiple different queries. Each query defines a category and asks for all the articles belonging to it and its subcategories. The results are shown in Table~\ref{tab:performances}. In the case of broad categories, the number of articles grows rapidly as we go deeper in the subcategory hierarchy. Each subcategory may have subcategories of its own and we define the depth to be the distance in hops from the initial category to the furthest subcategory in the subcategory tree.
For instance, the category \emph{Physics} already contains 33'917 articles and 972'206 hyperlinks at depth 4 and more than 2 million pages when articles are collected up to subcategory depth 7 as shown in Table~\ref{tab:physics_cat}. This is one-third of the articles of the English Wikipedia. This result is surprisingly large and additional investigation is required to understand the structure and check its correctness. The network is so connected that the number of links and the time to retrieve the data grows very quickly. After 4 hops in the category tree, it reaches tens of thousands of pages and more than a million links for some categories. 

This complexity in the category hierarchy makes it memory-expensive to query subgraphs of articles in the same category. Even though our solution slows down when the network expands, it is possible to query these subgraphs. Hence, our proposed database opens new avenues to the popularization of research on large sub-networks of categories. This may give a better understanding of the category and article structures. The results may lead to a better organization of categories and a more efficient process of verification of consistency in them.

\subsection{Combining the hyperlink graph and time series of visits}


It was shown that real-world events can be detected and tracked using Wikipedia viewership data. Besides, it is possible to use this data to detect abnormal patterns of visits in groups of connected Wikipedia articles~\cite{miz2019wikipedia}. For example, popular sports events such as Super Bowl, NBA playoffs, and FIFA World Cup, can be detected just by looking at Wikipedia viewership dynamics and the hyperlinks structure. Moreover, dramatic events such as airplane crashes or terrorist attacks can be spotted and analyzed. Also, the authors showed that it is possible to gain interesting insights related to the history of an event and its popularity among the users of the Web. This dataset will allow further investigations in this direction. 

Fig.~\ref{fig:wiki_snapshot} illustrates the result of the anomaly detection algorithm presented in~\cite{miz2019wikipedia}. Using the combination of the hyperlink network and the visitors' activity, the authors detected the groups of articles with simultaneous spikes in viewership dynamics. Besides, the authors used the dataset presented in this paper. They have published an interactive version of the results online as a part of the Wiki Insights project \footnote{\url{https://wiki-insights.epfl.ch/}}.


This approach allows selecting a subset of important hyperlinks out of the large amount present in each Wikipedia article. Indeed, following all the hyperlinks of some selected pages leads to very large graphs, as shown in Fig.~\ref{tab:perf_links}. New approaches are needed in order to create meaningful subgraphs by following only a reduced number of hyperlinks. This will help extract or emphasize particular types of information present in the network. In order to perform such studies, researchers can define different filters when querying the database we propose.


\section{Conclusion and future development}



In this paper, we presented a graph database to store and access Wikipedia web network and viewership activity of the pages. The main goal of this project was to provide a convenient tool for researchers working on Wikipedia and analyzing dynamic properties of this network. We designed the database with the idea of reproducible research in mind. We want this project to become an important building block in Wikipedia research community that should speed up the research process.


\subsection{Reproducible research}
In science, it should be mandatory for any published results to be reproducible. This implies an unlimited access to the data used for the experiments. However, when the dataset evolves with time, as it is in the case with Wikipedia articles and viewership statistics, it may be difficult to recover the exact data used in a given study. Some articles may have been removed or some links may have appeared after the publication of a scientific work. A database designed for scientists must include a mechanism for allowing experiments to be reproducible. One of the simplest solutions is as follows. The graph and time series are saved (frozen) every month. This allows the retrieval of graphs and pagecounts for any period in the past, while reducing the amount of memory required by the database.

\subsection{Potential benefits for Wikipedia}

A better understanding of Wikipedia structure, both from an article and a category point of view, is an important matter for the encyclopedia and the organization of its knowledge. Finding missing hyperlinks, suggesting links on pages creation, monitoring Wikipedia visitors activity, or structuring the category tree, are among the numerous possible applications.

\subsection{Enriching the database}

There are multiple ways for improving and enlarging the dataset. At the moment it contains only English Wikipedia. The number of languages can be extended. Articles with the same topic in different languages are naturally linked together in Wikipedia, which perfectly fits into the graph database framework. One could think of adding more data such as the text of articles for instance. To limit the graph database size, the best way would be to have a convenient toolkit that allows researchers to retrieve this information directly from Wikipedia dumps.

Information about Wikipedia edits and editors are also available. They could be structured as a graph of articles or a graph of users with time-series of edit activity, for example.

\balance

%
\begin{acks}
We thank the anonymous reviewers for their constructive feedback and useful recommendations that helped to improve this paper. V.~Miz has received funding from the European Union's H2020 Framework Programme (H2020-MSCA-ITN-2014) under grant agreement n\textsuperscript{o}~642685 MacSeNet.
\end{acks}

%
\bibliographystyle{ACM-Reference-Format}
\bibliography{biblio}

\end{document}